\documentclass[preprint,12pt,authoryear,fleqn]{elsarticle}
\usepackage[breaklinks=true, colorlinks=true]{hyperref}
\usepackage{microtype}
\usepackage{amsmath}
\usepackage[bitstream-charter]{mathdesign}
\usepackage[scaled=0.9]{courierten}
\usepackage[T1]{fontenc}

\journal{arXiv}

\begin{document}

\begin{frontmatter}


%% \affiliation{
%%     organization={},
%%     addressline={}, 
%%     city={},
%%     postcode={}, 
%%     state={},
%%     country={}

\title{Detecting temporal scaling with modified diffusion entropy analysis}

%% \affiliation[label1]{
%%     organization={},
%%     addressline={},
%%     city={},
%%     postcode={},
%%     state={},
%%     country={}
%% }
%% \affiliation[label2]{
%%     organization={},
%%     addressline={},
%%     city={},
%%     postcode={},
%%     state={},
%%     country={}
%% }

\author[inst1]{Garland Culbreth\fnref{label2}}
\fntext[label2]{Current affiliation: Institute for Health Metrics and Evaluation, University of Washington, 3980 15th Ave.
NE, Seattle, 98195, WA, USA.}
\author[inst1]{Jacob Baxley}
\author[inst1]{David Lambert}
%% \author[inst2]{Author Two}
%% \author[inst1,inst2]{Author Three}

\affiliation[inst1]{
    organization={Center for Nonlinear Science, University of North Texas},
    addressline={1155 Union Cir}, 
    city={Denton},
    postcode={76203}, 
    state={TX},
    country={USA}
}

\begin{abstract}
We present a modification to the diffusion entropy analysis method for detecting temporal scaling. Diffusion entropy analysis detects temporal scaling in a data set by converting a time-series into a diffusion trajectory and using the entropy of that trajectory to measure the temporal scaling in the data. We modify this by performing an event detection step to construct the diffusion trajectory. The new modified diffusion entropy analysis offers substantial improvements over the original method, especially for noisy data. We describe the method's purpose, how it works step-by-step, its application, and future development.
\end{abstract}


% \begin{highlights}
% \end{highlights}


\end{frontmatter}


\section*{Introduction}

Diffusion Entropy Analysis (DEA) is a time-series analysis method for detecting temporal complexity in a dataset; such as heartbeat rhythm \citep{bohara2017crucial, tuladhar2018meditation, jelinek2020diffusion}, seismographs \citep{mega2003power}, or financial markets \citep{cai2006diffusion}. DEA converts the time-series into a diffusion trajectory and uses the deviation of this diffusion from that of ordinary Brownian motion as a measure of the temporal complexity in the data. Several publications describe the method, mostly in broad strokes and only with words, but none yet have provided a step by step breakdown of what is happening with figures to illustrate.

This tutorial describes the modified DEA (MDEA) formally set forth by \cite{Culbreth2019}. This is sometimes referred to as DEA with stripes. Stripes refers to lines or thresholds overlaid atop the time-series data at regularly spaced intervals. An event is said to occur at any time index where the data series crosses one of these stripes. MDEA uses the asymmetric diffusion method described by \cite{GRIGOLINI2001}: events are taken to be positive, in our case $+1$, and are cumulatively summed to generate a diffusion trajectory based on the time-series. Once this diffusion trajectory is made a moving window method is applied, as in the original DEA \citep{Scafetta2002}. The window is moved along the diffusion trajectory and many small trajectories generated. Their displacements are binned in a histogram to create a probability density, which is then used to compute the Shannon entropy. Window length, $l$, is then increased and the process repeated to get many values of the entropy, $S(l)$, one for each window length. The end goal is to plot $S(l)$ vs. $ln(l)$ and to use the relation between them to extract the scaling, $\delta$, of the time-series process.

\section*{Method}

\begin{figure*}
    \centering
    \includegraphics[width=1.0\linewidth]{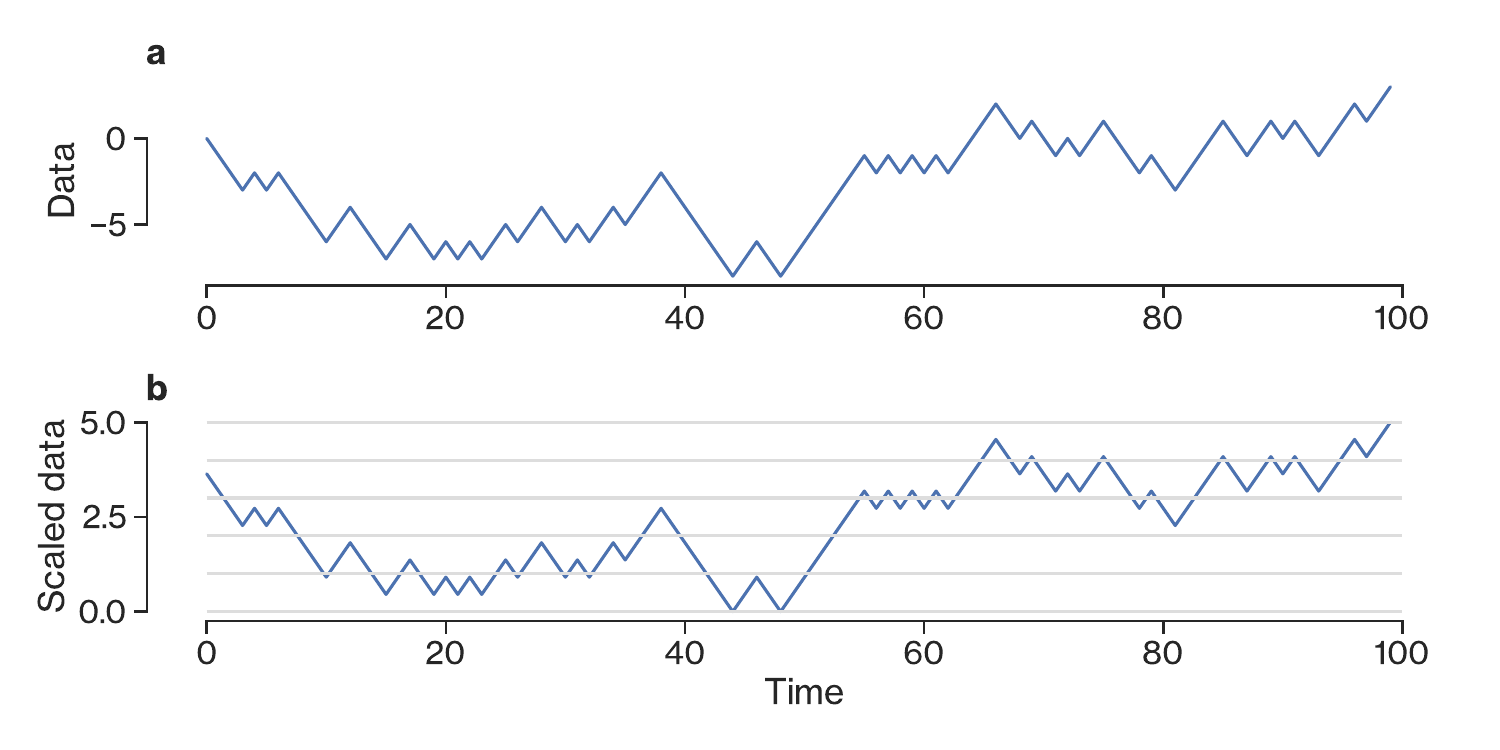}
    \caption{Original data and rounded data. (a) The original data series. (b) The corresponding rounded data series, shifted to fit within evenly spaced intervals according to the number of stripes chosen, with grid lines drawn at stripe boundaries.}
    \label{fig_data_and_stripes}
\end{figure*}

\begin{figure*}
    \centering
    \includegraphics[width=1.0\linewidth]{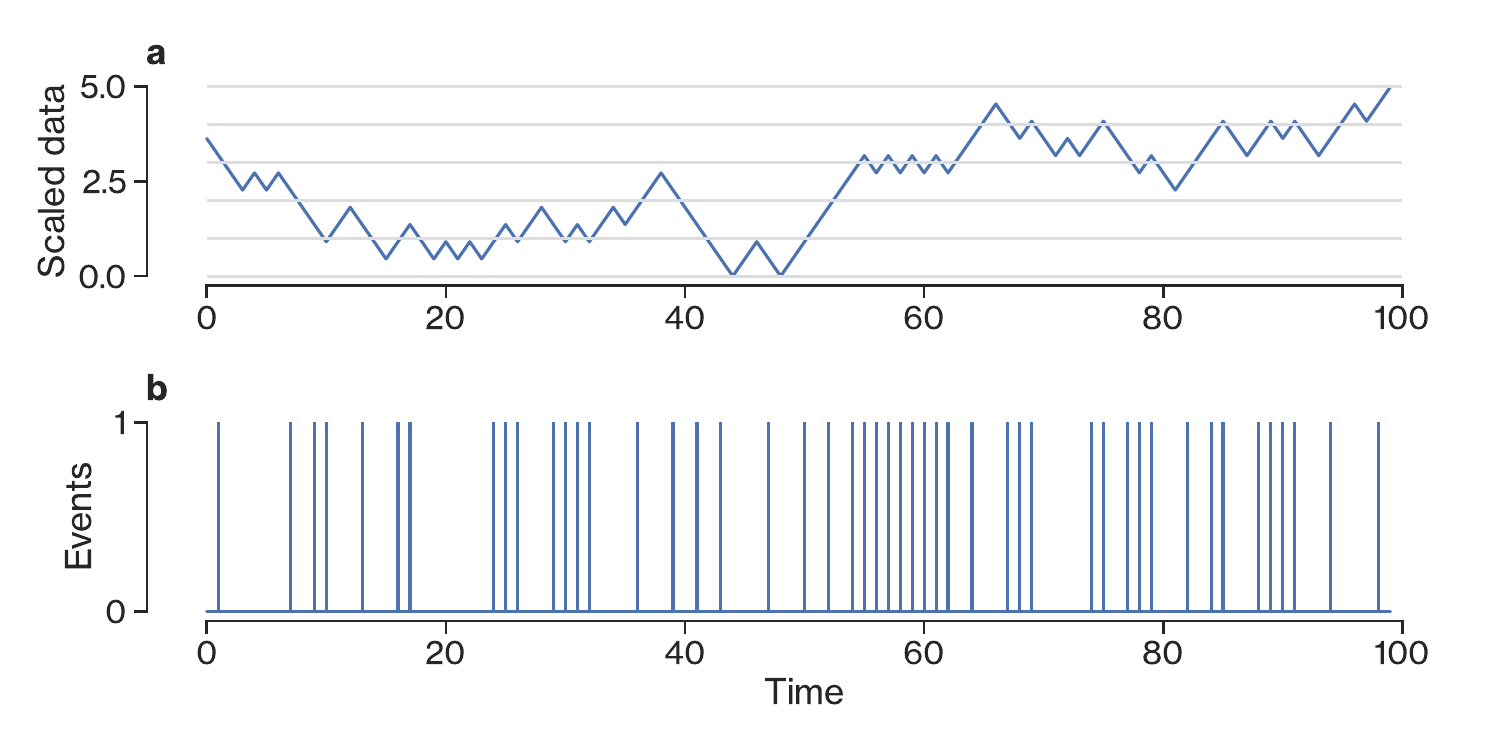}
    \caption{Rounded data and events. (a) The rounded data. (b) Events being recorded at every time index in the rounded data where either Equation \ref{eq:floor_check} or Equation \ref{eq:ceil_check} are false. All events have magnitude 1 and are positive \citep{GRIGOLINI2001}.}
    \label{fig_rounded_data_and_events}
\end{figure*}

\begin{figure*}
    \centering
    \includegraphics[width=1.0\linewidth]{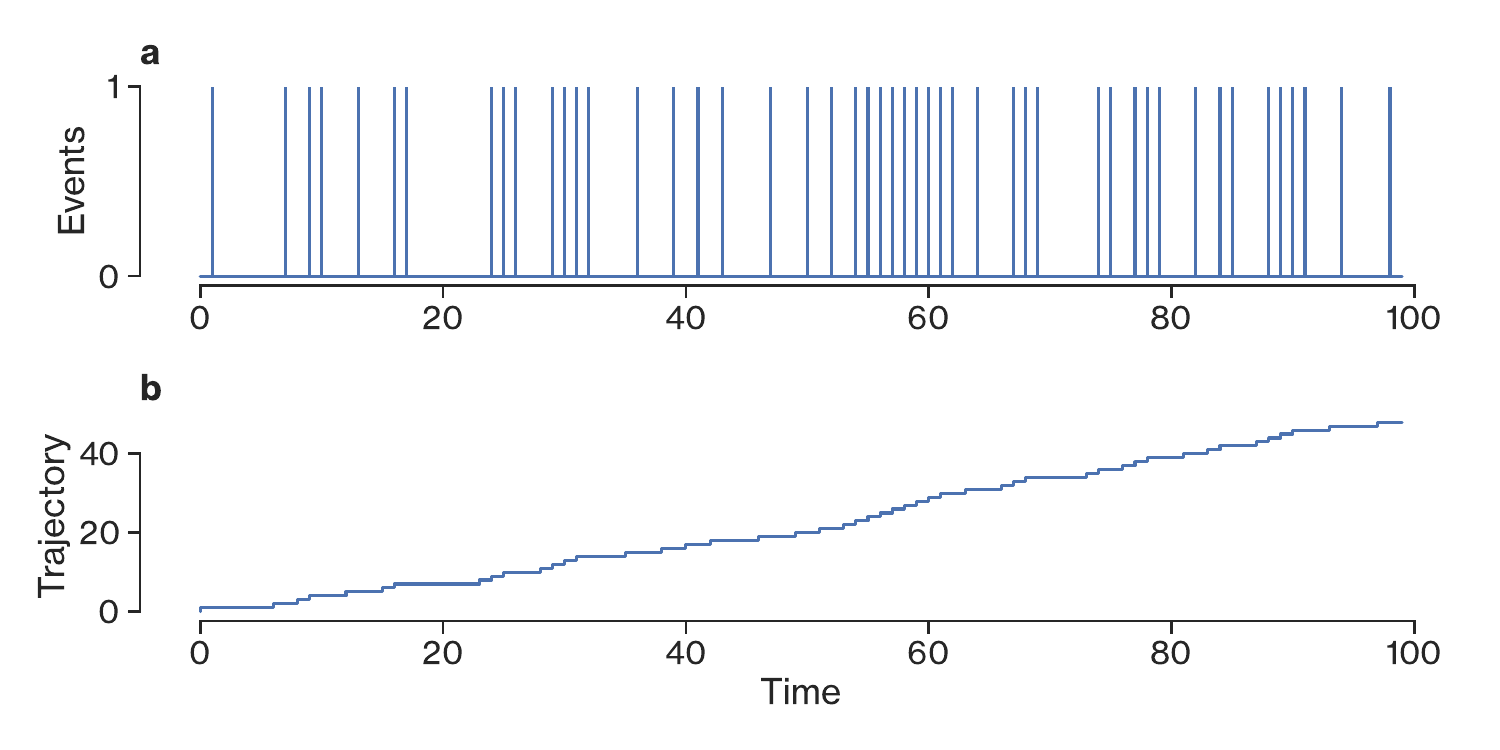}
    \caption{Events and diffusion trajectory. (a) Event array. (b) Diffusion trajectory constructed by taking cumulative sum of the events array.}
    \label{fig_events_and_trajectory}
\end{figure*}

\begin{figure}
    \centering
    \includegraphics[trim=0cm 0cm 0cm 0cm, clip, width=1.0\textwidth]{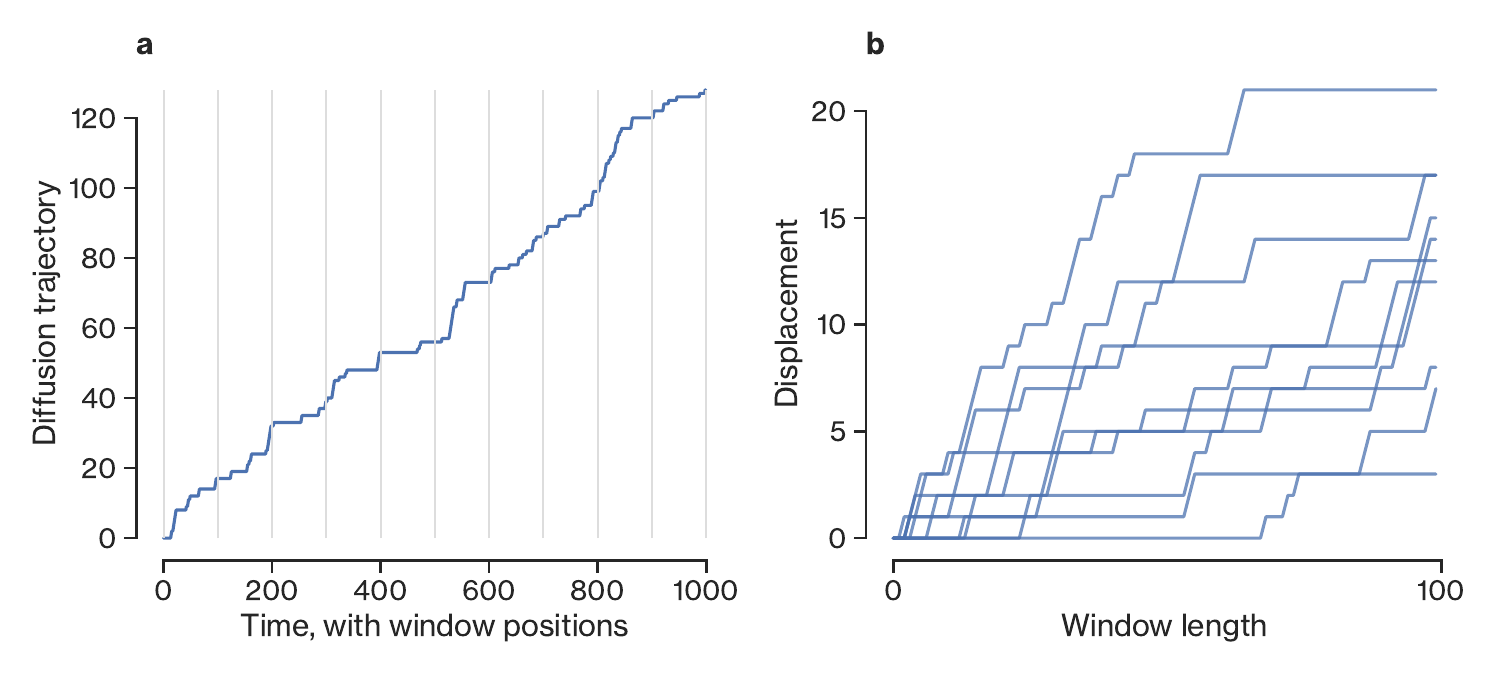}
    \caption{Moving window and trajectory slices. (a) The diffusion trajectory with grid lines drawn at a set of window positions, using $l = 100$. (b) The slices of the diffusion trajectory across each of the window positions in (a), laid atop each other and shifted to start from the same origin along the $y$-axis. Here the first 1000 steps in the series are used.}
    \label{fig_windows_and_slices}
\end{figure}

\begin{figure}
    \centering
    \includegraphics[width=0.5\linewidth]{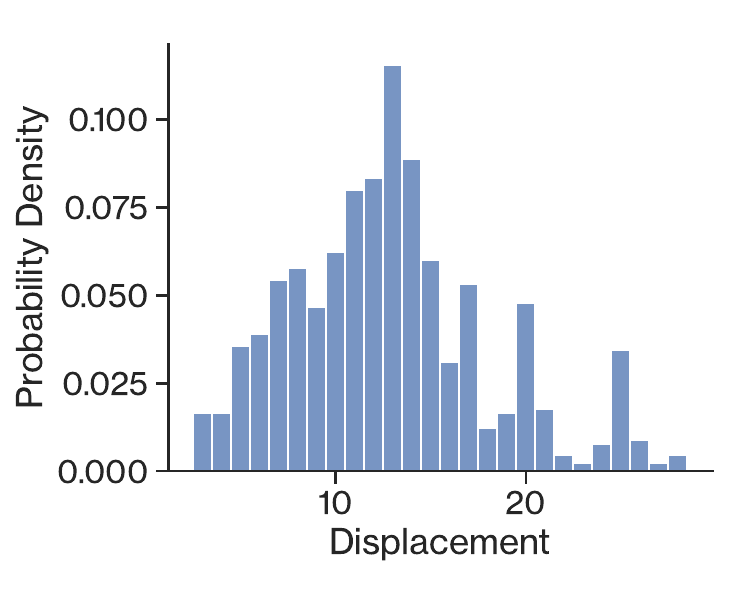}
    \caption{Distribution of diffusion slices. Histogram of the displacements of diffusion trajectory slices; normalized to form a probability distribution, $P(l)$.}
    \label{fig_distribution_histogram}
\end{figure}

\begin{figure}
    \centering
    \includegraphics[trim = 0cm 0cm 0cm 0cm, clip, width=1.0\linewidth]{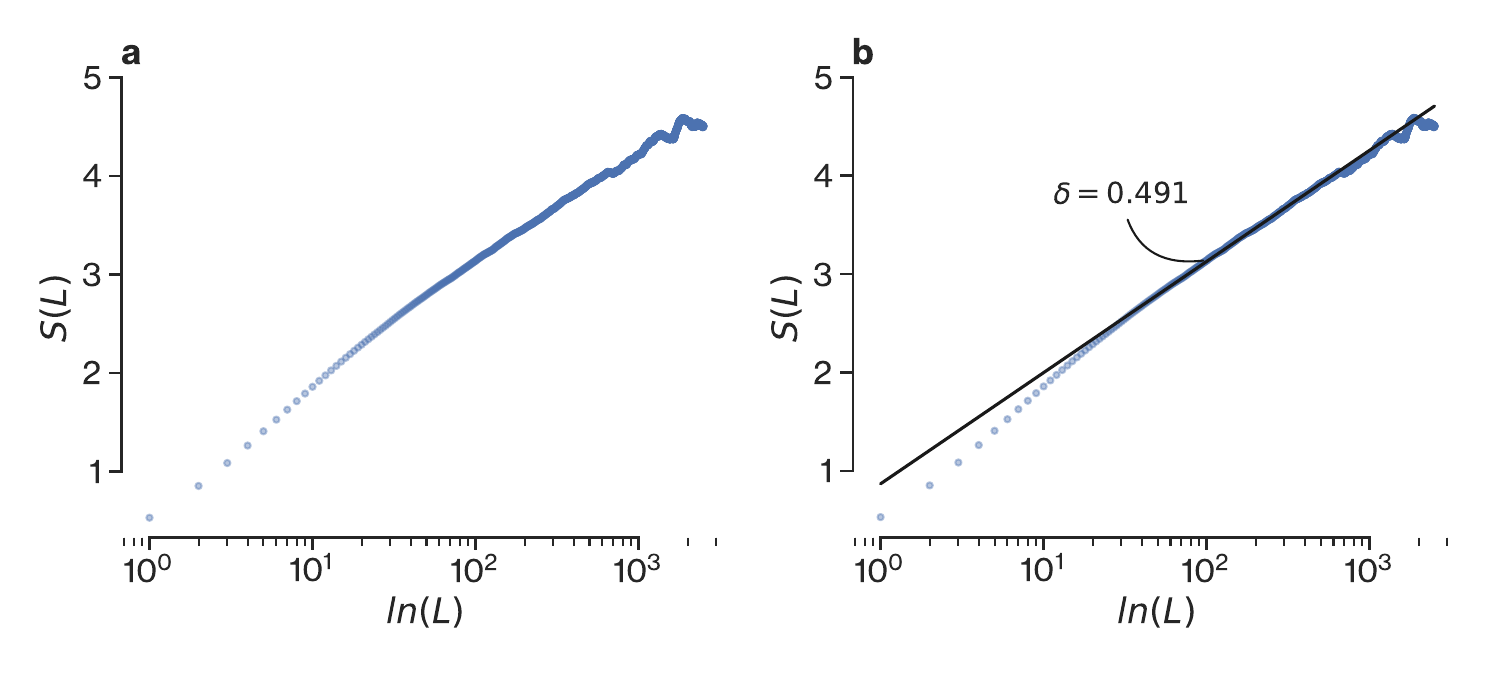}
    \caption{Principal result of MDEA. (a) Shannon entropy $S(l)$ vs. natural log of window lengths $\ln(l)$. The choice of natural log of $l$ for axis scale reveals any extant power-law behavior. (b) The slope of the linear fit, $\delta$, is the scaling of the time-series process. For a random walk such as this data, $\delta = 0.5$ is the expected value for MDEA to measure.}
    \label{fig:entropy}
\end{figure}

To demonstrate the method, a random walk of 10,000 steps is put into the MDEA program for analysis. Each step in the process is described in words and illustrated with a figure. Since the full 10,000 step series is difficult to visually digest, Figures \ref{fig_data_and_stripes}--\ref{fig_events_and_trajectory} display only the first 100 steps and Figures \ref{fig_windows_and_slices}--\ref{fig_distribution_histogram} the first 1,000 steps. Figure \ref{fig:entropy} shows the result of MDEA when run over the full 10,000 step series.

% \subsection{Stripes}

The first step in MDEA is to apply stripes to the data series under analysis. The researcher must define a number of stripes to use. Rigorous rules for this choice have yet to be formulated; a good practice is to choose a number between 2 and 100. The stripes are applied to the data by finding the maximum and minimum of the data, $x$, and computing the width, $w_\mathrm{x}$:
\begin{equation}
    \mathit{w_\mathrm{x}} = \left| \mathrm{max}(\mathit{x}) - \mathrm{min}(x) \right| .
\end{equation} 
The size of the stripes, $w_\mathrm{s}$, on the data width, $w_\mathrm{x}$, are then computed by taking:
\begin{equation}
    \mathit{w_\mathrm{s}} = \frac{w_\mathrm{x}}{\mathit{N}},
\end{equation}
where $N$ is the number of stripes to be applied. The data is then uniformly scaled so that the stripes have width 1 according to:
\begin{equation}
    \mathit{X} = \frac{x}{\mathit{w_\mathrm{s}}}.
\end{equation}
The result of this process is shown in Figure \ref{fig_data_and_stripes}. 

% \subsection{Events and Diffusion}

Once the data has been rounded to the stripe widths, events are recorded at all times when the data crosses, or intersects, a stripe. To determine whether a crossing occurs at time $t$, MDEA checks two conditions:
\begin{equation} \label{eq:floor_check}
    X_t \stackrel{?}{<} \mathrm{floor} (X_{t-1} + 1) ,
\end{equation}
\begin{equation} \label{eq:ceil_check}
    X_t \stackrel{?}{>} \mathrm{ceil} (X_{t-1} - 1) .
\end{equation}
If either of these conditions is false then a crossing must have occurred at time $t$. If a crossing occurs at time $t$ a 1 is appended to the event array, $E$, at index $t$; if not a 0 is appended. This process is shown in Figure \ref{fig_rounded_data_and_events}. 

The next step is to construct a diffusion trajectory, $Y$, from the recorded events. This is done by taking the cumulative sum of the event array, $E$:
\begin{equation}
    Y_n = \sum_{t=0}^n E_t,
\end{equation}
where $n$ starts off at the first index of the array---0 in most programming languages---then increases by 1 until $n$ equals the length of the array. This is depicted in Figure \ref{fig_events_and_trajectory}.

% \subsection{Moving Window}

DEA uses a moving window to examine this diffusion trajectory. The process is similar to some other time-series analysis methods which use moving windows. In particular, the moving window is used similarly to Detrended Fluctuation Analysis (DFA) \citep{peng1994mosaic}. The difference is that DFA uses windows that do not overlap, while DEA uses window positions advancing one time index by one with much overlap. This similarity and the rationale behind this decision is discussed in \cite{GRIGOLINI2001}. For this demonstration only one window length, $l=100$, is defined, which is placed at fixed positions (end to end) on the diffusion trajectory. This makes the windows and snapshots easier to visualize. In practice a set of window lengths is defined and each stepped down the trajectory, shifting one index at a time. For this demonstration only one possible window length is used and only a subset of the window positions.

Once these trajectory slices have been computed the next step is to make a histogram of their displacements. For each window length $l$, stepping the window along the trajectory generates many slices, each of which goes some distance from the origin -- along the y-axis in Figure \ref{fig_windows_and_slices}. These displacements are binned in a histogram, which is then normalized to turn it into a probability distribution, $P(l)$. One such histogram is plotted in Figure \ref{fig_distribution_histogram}.

% \subsection{Entropy and Scaling}

Information entropy, or Shannon entropy \citep{Shannon1948}, is defined:
\begin{equation}
    S(Y) = - \sum_i P(y_i) \log(P(y_i)) ,
\end{equation}
where $S$ is the entropy, $Y$ the random variable, $y_i$ the possible values of $Y$, and $P(y_i)$ the probabilities associated with each value. For DEA, with $Y$ being the value of the diffusion trajectory at $t = l$ and with probability distributions $P_l$ corresponding to window length $l$, this goes:
\begin{equation} \label{eq:dea_entropy}
    S_l(Y) = - \sum_i P_l(y_i) \ln(P_l(y_i)) . 
\end{equation} 
DEA uses many window lengths, constructs probability distributions for each, and calculates the entropy corresponding to each according to Equation \ref{eq:dea_entropy}. Figure \ref{fig:entropy} illustrates an example of the result. 

The goal of DEA is to extract the scaling of the data series. This is done by using the relation between $S$ and $l$:
\begin{equation} \label{eq:dea_scaling}
    S(l) = \delta\ln(l) + C,
\end{equation}
where $\delta$ is the scaling of the process and $C$ is some constant \citep{Scafetta2002}. This relation is linear, in log-scale. Therefore, if a linear fit is performed in log-scale, as is shown in Figure \ref{fig:entropy}, the slope is the scaling, $\delta$, of the time-series.

\citet{clauset2009power} show that using least-squares to fit the slope for a line of the form in equation \ref{eq:dea_scaling} introduces systematic errors. MDEA therefore includes support for using more robust linear fitting: the \citet{siegel1982robust} and \citet{theil1950rank}-\citet{sen1968estimates} methods. These methods fit an ensemble of lines between pairs of data points and use the median slope of the ensemble as the slope of the fit line. This makes them much more robust against outliers, and avoids the systematic errors introduced by least-squares.
%% GC: The Clauset 2009 paper gives in appendix A an equation with exactly the same form as the one we use for S(l). That's where they comment that using least-squares to estimate the slope of a line with that form introduces systematic error, and since the equation is the same and their statement isn't confined to fitting probability densities, I'm back to not trusting least-squares for DEA. However, the Theil-Sen and Siegel methods seem to avoid that problem, so I implemented them into the DEA code on github, with Seigel being set to the default fit method.

\section*{Discussion}

MDEA measures a kind of temporal complexity related to criticality. The scaling, $\delta$, that MDEA outputs provides a measure of the complexity present in the data set. The behavior of $\delta$ and its relation to criticality and complexity parameters is discussed in depth in \cite{GRIGOLINI2001} and \cite{Scafetta2002}. For this tutorial it suffices to understand the basic rules for interpreting the results. For a completely non-complex process, such as a random walk, MDEA yields $\delta = 0.5$. For a process at criticality, MDEA yields $\delta = 1$. Therefore, $\delta$ represents a measure of the complexity present in the process: the closer $\delta$ is to 1 the closer the process is to criticality.

MDEA offers significant advantages over the original DEA. In DEA the original time-series is used as the diffusion trajectory with no pre-processing. As a result several kinds of noise are able to affect the measured scaling. For example, if the time-series is characterized by fractional Brownian motion (FBM) \citep{Mandelbrot1968}, DEA will measure a scaling equal to the Hurst index of the FBM. Even in cases where FBM is only a component of the data and not a complete characterization, the scaling measured by DEA is significantly affected by the FBM contribution. MDEA filters out any existing FBM contribution to the scaling, allowing examination and quantification of other complexity sources in the data \citep{Culbreth2019}.

In particular, MDEA improves detection of events that occur at time intervals distributed according to a power law. These events appear in several complex processes, and they are conjectured to be a signature of a complex system undergoing self-organization in time \citep{bohara2018bridging}.

\section*{Future Development}

MDEA development is ongoing and several key areas remain where MDEA requires improvement.\footnote{As of \today, the MDEA code repository is located at: \url{https://github.com/garland-culbreth/Diffusion-Entropy-Analysis}} Most urgent of these is the determination of fitting interval. So far this interval has been chosen on a case-by-case basis by the researcher looking at the plot of $S(l)$ vs. $\ln(l)$ and adjusting the fit to lie along the linear portion. This is not rigorous and quite unsatisfactory. Using the \citet{siegel1982robust} and \citet{theil1950rank}-\citet{sen1968estimates} robust linear fitting methods removes the need to set a lower bound for this fitting interval, but still require an upper bound. A rigorous and standardized method of determining the correct fitting interval is urgently needed, perhaps by performing iterated fits and adjusting the fit interval based on the residuals. Alternatively, since equation \ref{eq:dea_scaling} has a power law form, it may be possible to use maximum likelihood estimation to obtain the scaling \citep{clauset2009power}. Using maximum likelihood estimation would remove the need to set a fit interval at all.

Secondly, the method for determining how many stripes ought to be used requires improvement. So far this is another case-by-case choice on the part of the researcher, usually done by testing several possible values and trying to guess which one is best. The best guidance thus far developed is that the correct number of stripes is usually between 2 and 70. However, there is currently no rigorous method for determining the correct number, or even for judging whether the researcher's guess is good or bad. Further work is urgently needed to develop a rigorous method for determining how many stripes ought to be used, which should follow the direction established by \citet{allegrini2002real}.
%If we want to talk about this, there was some work by Paolo Allegrini and others around 2002 or so about this. -David
% I think I remember reading that paper, or one that mentioned work on this, a while ago (a year or more?). I'll see if I can find the reference to put it in here and mention their work. -GC

Finally, the MDEA code requires further optimization, especially for run-time speed. The MDEA code hosted on GitHub is written in Python and uses NumPy \citep{harris2020array}, which supports using Fortran code modules to speed up computations. Future development should investigate whether this functionality can be implemented into MDEA, or should develop a bespoke integration using modules written in other high-performance languages, such as Rust. Development in this direction would provide substantial runtime improvements.

\bibliographystyle{elsarticle-harv} 
\bibliography{references}

\begin{thebibliography}{18}
\expandafter\ifx\csname natexlab\endcsname\relax\def\natexlab#1{#1}\fi
\providecommand{\url}[1]{\texttt{#1}}
\providecommand{\href}[2]{#2}
\providecommand{\path}[1]{#1}
\providecommand{\DOIprefix}{doi:}
\providecommand{\ArXivprefix}{arXiv:}
\providecommand{\URLprefix}{URL: }
\providecommand{\Pubmedprefix}{pmid:}
\providecommand{\doi}[1]{\href{http://dx.doi.org/#1}{\path{#1}}}
\providecommand{\Pubmed}[1]{\href{pmid:#1}{\path{#1}}}
\providecommand{\bibinfo}[2]{#2}
\ifx\xfnm\relax \def\xfnm[#1]{\unskip,\space#1}\fi
%Type = Article
\bibitem[{Allegrini et~al.(2002)Allegrini, Balocchi, Chillemi, Grigolini,
  Hamilton, Maestri, Palatella and Raffaelli}]{allegrini2002real}
\bibinfo{author}{Allegrini, P.}, \bibinfo{author}{Balocchi, R.},
  \bibinfo{author}{Chillemi, S.}, \bibinfo{author}{Grigolini, P.},
  \bibinfo{author}{Hamilton, P.}, \bibinfo{author}{Maestri, R.},
  \bibinfo{author}{Palatella, L.}, \bibinfo{author}{Raffaelli, G.},
  \bibinfo{year}{2002}.
\newblock \bibinfo{title}{Real event detection and the treatment of congestive
  heart failure: an efficient technique to help cardiologists to make crucial
  decisions}.
\newblock \bibinfo{journal}{arXiv preprint cond-mat/0209038} .
%Type = Article
\bibitem[{Bohara et~al.(2017)Bohara, Lambert, West and
  Grigolini}]{bohara2017crucial}
\bibinfo{author}{Bohara, G.}, \bibinfo{author}{Lambert, D.},
  \bibinfo{author}{West, B.J.}, \bibinfo{author}{Grigolini, P.},
  \bibinfo{year}{2017}.
\newblock \bibinfo{title}{Crucial events, randomness, and multifractality in
  heartbeats}.
\newblock \bibinfo{journal}{Physical Review E} \bibinfo{volume}{96},
  \bibinfo{pages}{062216}.
%Type = Article
\bibitem[{Bohara et~al.(2018)Bohara, West and Grigolini}]{bohara2018bridging}
\bibinfo{author}{Bohara, G.}, \bibinfo{author}{West, B.J.},
  \bibinfo{author}{Grigolini, P.}, \bibinfo{year}{2018}.
\newblock \bibinfo{title}{Bridging waves and crucial events in the dynamics of
  the brain}.
\newblock \bibinfo{journal}{Frontiers in physiology} \bibinfo{volume}{9},
  \bibinfo{pages}{1174}.
%Type = Article
\bibitem[{Cai et~al.(2006)Cai, Zhou, Yang, Yang, Wang and
  Zhou}]{cai2006diffusion}
\bibinfo{author}{Cai, S.M.}, \bibinfo{author}{Zhou, P.L.},
  \bibinfo{author}{Yang, H.J.}, \bibinfo{author}{Yang, C.X.},
  \bibinfo{author}{Wang, B.H.}, \bibinfo{author}{Zhou, T.},
  \bibinfo{year}{2006}.
\newblock \bibinfo{title}{Diffusion entropy analysis on the scaling behavior of
  financial markets}.
\newblock \bibinfo{journal}{Physica A: Statistical Mechanics and its
  Applications} \bibinfo{volume}{367}, \bibinfo{pages}{337--344}.
%Type = Article
\bibitem[{Clauset et~al.(2009)Clauset, Shalizi and Newman}]{clauset2009power}
\bibinfo{author}{Clauset, A.}, \bibinfo{author}{Shalizi, C.R.},
  \bibinfo{author}{Newman, M.E.}, \bibinfo{year}{2009}.
\newblock \bibinfo{title}{Power-law distributions in empirical data}.
\newblock \bibinfo{journal}{SIAM review} \bibinfo{volume}{51},
  \bibinfo{pages}{661--703}.
%Type = Article
\bibitem[{Culbreth et~al.(2019)Culbreth, West and Grigolini}]{Culbreth2019}
\bibinfo{author}{Culbreth, G.}, \bibinfo{author}{West, B.},
  \bibinfo{author}{Grigolini, P.}, \bibinfo{year}{2019}.
\newblock \bibinfo{title}{Entropic approach to the detection of crucial
  events}.
\newblock \bibinfo{journal}{Entropy} \bibinfo{volume}{21},
  \bibinfo{pages}{178}.
\newblock \URLprefix \url{https://doi.org/10.3390/e21020178},
  \DOIprefix\doi{10.3390/e21020178}.
%Type = Article
\bibitem[{Grigolini et~al.(2001)Grigolini, Palatella and
  Raffelli}]{GRIGOLINI2001}
\bibinfo{author}{Grigolini, P.}, \bibinfo{author}{Palatella, L.},
  \bibinfo{author}{Raffelli, G.}, \bibinfo{year}{2001}.
\newblock \bibinfo{title}{Asymmetric anomalous diffusion: an efficient way to
  detect memory in time series}.
\newblock \bibinfo{journal}{Fractals} \bibinfo{volume}{09},
  \bibinfo{pages}{439--449}.
\newblock \URLprefix \url{https://doi.org/10.1142/s0218348x01000865},
  \DOIprefix\doi{10.1142/s0218348x01000865}.
%Type = Article
\bibitem[{Harris et~al.(2020)Harris, Millman, van~der Walt, Gommers, Virtanen,
  Cournapeau, Wieser, Taylor, Berg, Smith et~al.}]{harris2020array}
\bibinfo{author}{Harris, C.R.}, \bibinfo{author}{Millman, K.J.},
  \bibinfo{author}{van~der Walt, S.J.}, \bibinfo{author}{Gommers, R.},
  \bibinfo{author}{Virtanen, P.}, \bibinfo{author}{Cournapeau, D.},
  \bibinfo{author}{Wieser, E.}, \bibinfo{author}{Taylor, J.},
  \bibinfo{author}{Berg, S.}, \bibinfo{author}{Smith, N.J.}, et~al.,
  \bibinfo{year}{2020}.
\newblock \bibinfo{title}{Array programming with numpy}.
\newblock \bibinfo{journal}{Nature} \bibinfo{volume}{585},
  \bibinfo{pages}{357--362}.
%Type = Article
\bibitem[{Jelinek et~al.(2020)Jelinek, Tuladhar, Culbreth, Bohara, Cornforth,
  West and Grigolini}]{jelinek2020diffusion}
\bibinfo{author}{Jelinek, H.F.}, \bibinfo{author}{Tuladhar, R.},
  \bibinfo{author}{Culbreth, G.}, \bibinfo{author}{Bohara, G.},
  \bibinfo{author}{Cornforth, D.}, \bibinfo{author}{West, B.J.},
  \bibinfo{author}{Grigolini, P.}, \bibinfo{year}{2020}.
\newblock \bibinfo{title}{Diffusion entropy vs. multiscale and r{\'e}nyi
  entropy to detect progression of autonomic neuropathy}.
\newblock \bibinfo{journal}{Frontiers in Physiology} \bibinfo{volume}{11}.
%Type = Article
\bibitem[{Mandelbrot and Ness(1968)}]{Mandelbrot1968}
\bibinfo{author}{Mandelbrot, B.B.}, \bibinfo{author}{Ness, J.W.V.},
  \bibinfo{year}{1968}.
\newblock \bibinfo{title}{Fractional brownian motions, fractional noises and
  applications}.
\newblock \bibinfo{journal}{{SIAM} Review} \bibinfo{volume}{10},
  \bibinfo{pages}{422--437}.
\newblock \URLprefix \url{https://doi.org/10.1137/1010093},
  \DOIprefix\doi{10.1137/1010093}.
%Type = Article
\bibitem[{Mega et~al.(2003)Mega, Allegrini, Grigolini, Latora, Palatella,
  Rapisarda and Vinciguerra}]{mega2003power}
\bibinfo{author}{Mega, M.S.}, \bibinfo{author}{Allegrini, P.},
  \bibinfo{author}{Grigolini, P.}, \bibinfo{author}{Latora, V.},
  \bibinfo{author}{Palatella, L.}, \bibinfo{author}{Rapisarda, A.},
  \bibinfo{author}{Vinciguerra, S.}, \bibinfo{year}{2003}.
\newblock \bibinfo{title}{Power-law time distribution of large earthquakes}.
\newblock \bibinfo{journal}{Physical Review Letters} \bibinfo{volume}{90},
  \bibinfo{pages}{188501}.
%Type = Article
\bibitem[{Peng et~al.(1994)Peng, Buldyrev, Havlin, Simons, Stanley and
  Goldberger}]{peng1994mosaic}
\bibinfo{author}{Peng, C.K.}, \bibinfo{author}{Buldyrev, S.V.},
  \bibinfo{author}{Havlin, S.}, \bibinfo{author}{Simons, M.},
  \bibinfo{author}{Stanley, H.E.}, \bibinfo{author}{Goldberger, A.L.},
  \bibinfo{year}{1994}.
\newblock \bibinfo{title}{Mosaic organization of dna nucleotides}.
\newblock \bibinfo{journal}{Physical review e} \bibinfo{volume}{49},
  \bibinfo{pages}{1685}.
%Type = Article
\bibitem[{Scafetta and Grigolini(2002)}]{Scafetta2002}
\bibinfo{author}{Scafetta, N.}, \bibinfo{author}{Grigolini, P.},
  \bibinfo{year}{2002}.
\newblock \bibinfo{title}{Scaling detection in time series: Diffusion entropy
  analysis}.
\newblock \bibinfo{journal}{Phys. Rev. E} \bibinfo{volume}{66},
  \bibinfo{pages}{036130}.
\newblock \URLprefix \url{https://link.aps.org/doi/10.1103/PhysRevE.66.036130},
  \DOIprefix\doi{10.1103/PhysRevE.66.036130}.
%Type = Article
\bibitem[{Sen(1968)}]{sen1968estimates}
\bibinfo{author}{Sen, P.K.}, \bibinfo{year}{1968}.
\newblock \bibinfo{title}{Estimates of the regression coefficient based on
  kendall's tau}.
\newblock \bibinfo{journal}{Journal of the American statistical association}
  \bibinfo{volume}{63}, \bibinfo{pages}{1379--1389}.
%Type = Article
\bibitem[{Shannon(1948)}]{Shannon1948}
\bibinfo{author}{Shannon, C.E.}, \bibinfo{year}{1948}.
\newblock \bibinfo{title}{A mathematical theory of communication}.
\newblock \bibinfo{journal}{Bell System Technical Journal}
  \bibinfo{volume}{27}, \bibinfo{pages}{379--423}.
\newblock \URLprefix \url{https://doi.org/10.1002/j.1538-7305.1948.tb01338.x},
  \DOIprefix\doi{10.1002/j.1538-7305.1948.tb01338.x}.
%Type = Article
\bibitem[{Siegel(1982)}]{siegel1982robust}
\bibinfo{author}{Siegel, A.F.}, \bibinfo{year}{1982}.
\newblock \bibinfo{title}{Robust regression using repeated medians}.
\newblock \bibinfo{journal}{Biometrika} \bibinfo{volume}{69},
  \bibinfo{pages}{242--244}.
%Type = Article
\bibitem[{Theil(1950)}]{theil1950rank}
\bibinfo{author}{Theil, H.}, \bibinfo{year}{1950}.
\newblock \bibinfo{title}{A rank-invariant method of linear and polynomial
  regression analysis}.
\newblock \bibinfo{journal}{Indagationes mathematicae} \bibinfo{volume}{12},
  \bibinfo{pages}{173}.
%Type = Article
\bibitem[{Tuladhar et~al.(2018)Tuladhar, Bohara, Grigolini and
  West}]{tuladhar2018meditation}
\bibinfo{author}{Tuladhar, R.}, \bibinfo{author}{Bohara, G.},
  \bibinfo{author}{Grigolini, P.}, \bibinfo{author}{West, B.J.},
  \bibinfo{year}{2018}.
\newblock \bibinfo{title}{Meditation-induced coherence and crucial events}.
\newblock \bibinfo{journal}{Frontiers in physiology} \bibinfo{volume}{9},
  \bibinfo{pages}{626}.

\end{thebibliography}

\end{document}